\documentclass[11pt]{article}
\usepackage[T1]{fontenc}
\usepackage[utf8]{inputenc}
\usepackage{amsfonts}
\usepackage{amsmath}
\usepackage{enumerate}
\usepackage{amssymb}
\usepackage{amsthm}
\usepackage{bbm}
\usepackage{bm}
\usepackage{mathrsfs}
\usepackage{verbatim}
\usepackage{setspace}
\usepackage{color}
\usepackage{pdfsync}
\usepackage{booktabs}
\usepackage{graphicx}
\usepackage{amsmath,varwidth,array,ragged2e}
\usepackage{tabu}
\usepackage{array}
\usepackage{makecell}

\usepackage{enumitem}
\theoremstyle{plain}
\newtheorem{theorem}{Theorem}[section]

\newtheorem{corollary}[theorem]{Corollary}

\theoremstyle{definition}
\newtheorem{definition}[theorem]{Definition}
\newtheorem{remark}[theorem]{Remark}

\newtheorem{example}[theorem]{Example}

\theoremstyle{remark}

\renewenvironment{thebibliography}[1]{%
\begin{oldthebibliography}{#1}%
\setlength{\baselineskip}{.9em}
\linespread{1}
\small
\setlength{\parskip}{0.3ex}%
\setlength{\itemsep}{.5em}%
}%
{%
\end{oldthebibliography}%
}

\DeclareMathOperator{\conv}{conv}

\numberwithin{equation}{section}

\usepackage[pdfborder={0 0 0}]{hyperref}
\hypersetup{
  urlcolor = black,
  pdfauthor = {Stephan Eckstein, Michael Kupper},
  pdfkeywords = {T.B.D.;
},
}

\begin{document}

\title{Martingale transport with homogeneous stock movements}
\author{Stephan Eckstein\thanks{Department of Mathematics, University of Konstanz, Universit\"{a}tsstra{\ss}e 10, 78464 Konstanz, Germany, stephan.eckstein@uni-konstanz.de} \and Michael Kupper\thanks{Department of Mathematics, University of Konstanz, Universit\"{a}tsstra{\ss}e 10, 78464 Konstanz, Germany, kupper@uni-konstanz.de} 
}
\date{\today}

\maketitle

\begin{abstract}
	We study a variant of the martingale optimal transport problem in a multi-period setting to derive robust price bounds of a financial derivative. On top of marginal and martingale constraints, we introduce a time-homogeneity assumption, which restricts the variability of the forward-looking transitions of the martingale across time. We provide a dual formulation in terms of superhedging and discuss relaxations of the time-homogeneity assumption by adding market frictions. In financial terms, the introduced time-homogeneity corresponds to a time-consistency condition for call prices, given the state of the stock. The time homogeneity assumption leads to improved price bounds as market data from many time points can be incorporated effectively. The approach is illustrated with two numerical examples.
\end{abstract}

\section{Introduction}
We consider a discrete stock process $S_1, \ldots, S_T$. The goal is to find a fair price of a financial instrument $f(S_1, \ldots, S_T)$ depending on this stock. We follow the robust pricing idea of martingale optimal transport \cite{beiglbock2013model,beiglbock2016problem}, in that we determine the highest and lowest possible price for this instrument under pricing rules which are consistent with European call and put prices observed on the market (which determine the risk-neutral one-period marginal distributions of $S_1, \ldots, S_T$) and the assumption that the process $S_1, \ldots, S_T$ is a martingale. In addition, this paper adds a notion of time-homogeneity for the process $S_1, \ldots, S_T$, made precise in Section \ref{sec:homogeneity}. The reason we introduce this assumption is twofold:
\begin{itemize}
	\item[(1)] While the martingale optimal transport approach is very robust, for practical purposes the obtained range of prices is often too wide, see \cite{henry2013automated,lutkebohmert2019tightening,sester2018markov}.
	\item[(2)] In martingale optimal transport, the transition probabilities of the considered martingale models are almost entirely decoupled. For instance the transition probabilities from period 1 to 2, and 3 to 4, can be completely different. Hence market information which restricts the possible transitions from period 1 to 2 has practically no relevance when pricing an instrument depending only on time points 3 and 4 (for a numerical illustration, see Section \ref{subsec:binomial}).
\end{itemize}
The notion of time-homogeneity of the stock process we introduce mainly aims at putting the transition probabilities between different periods in relation and thus improve on the issue raised in point (2). This leads to a more narrow range of possible prices to improve on point (1).

To introduce the notion of time-homogeneity we use, let us recall homogeneous Markov models. Intuitively speaking, a homogeneous Markov model $(S_1, \ldots, S_T)$ satisfies the following two properties:
\begin{itemize}
	\item[(i)] The conditional distribution of $S_{t+1}$ given $(S_1, \ldots, S_t)$ equals that of $S_{t+1}$ given $S_t$.
	\item[(ii)] The conditional distribution of $S_{t+1}$ given $S_t$ equals that of $S_{s+1}$ given $S_s$.
\end{itemize}
While Markovian models in a martingale optimal transport framework have been considered in \cite{sester2018markov}, the assumption is difficult to handle both regarding duality, and numerics,  since the set of Markovian models is not convex.

To achieve the goal set out in this paper however, only the second property (ii) of homogeneous Markov models is required, as this is the property which couples the transition probabilities across different time periods. We hence say that a process is homogeneous if it only satisfies property (ii), made precise in Definition \ref{def:Qhom}. The intuition that every homogeneous Markov model is homogeneous also holds rigorously, which is established in Remark \ref{remark:basics} alongside other properties and characterisations of homogeneity.

Three key features of incorporating homogeneity into the martingale optimal transport setting are worth pointing out: First, the homogeneous martingale optimal transport problem is as numerically tractable as the martingale optimal transport problem without time-homogeneity, in that the discretized version reduces to a linear program and the dual formulation is well suited for various approaches, see, e.g.,~\cite{eckstein2018computation,guo2019computational,henry2013automated}. Second, the dual formulation can be interpreted in terms of trading strategies and superhedging. And third, market frictions and relaxations of the introduced time-homogeneity assumption can be incorporated naturally.

In the recent literature, different methods have been studied to improve on 
point (1) above and hence make the martingale optimal transport approach more 
practicable. In \cite{lutkebohmert2019tightening, 
	sester2018markov} the authors study additional variance and Markovianity 
constraints on the underlying stock 
process. In \cite{guyon2019joint} additional information from options written 
on the stock's volatility is incorporated.

The rest of the paper is structured as follows: In Section 2, we give the relevant notation and recall basic facts about martingale optimal transport. In Section 3, the notion of time-homogeneity is introduced and we state basic properties and duality for the time-homogeneous version of the martingale optimal transport problem. In Section 4, extensions like market frictions, relaxed assumptions and higher dimensional markets are discussed. Section 5 gives two short numerical examples. All proofs are postponed to Section 6. The Appendix discusses the technical assumption $(A)$ which is made to obtain the main Theorem \ref{thm:duality}.

\section{Notation and Martingale Optimal Transport}
$S = (S_1, \ldots, S_T)$ denotes the value of a stock at time points $t=1, \ldots, T$, which model an equally spaced time-grid. For simplicity, we assume no risk-free rate and no dividends. We model the asset prices as the canonical process on $\mathbb{R}^T$, i.e.,~$S_t(\omega) = \omega_t$ for $\omega \in \mathbb{R}^T$. Here, $\mathbb{R}^T$ is endowed with the Borel $\sigma$-algebra $\mathcal{B}(\mathbb{R}^T)$ and Euclidean norm $|\cdot|$. We denote by $C_{\rm lin}(\mathbb{R}^T)$ (resp.~$C_b(\mathbb{R}^T)$) the set of all continuous functions $f\colon\mathbb{R}^T\to\mathbb{R}$ such that $|f(\cdot)|/(1+|\cdot|)$ is bounded (resp.~$f$ is bounded)
and by $\mathcal{P}(\mathbb{R}^T)$ the set of all probability measures $\mathbb{Q}$ on
$\mathcal{B}(\mathbb{R}^T)$.

Let $\mu_1, \ldots, \mu_T \in \mathcal{P}(\mathbb{R})$ have finite first moments. The measures $\mu_1, \ldots, \mu_T$ model the risk-neutral marginal distributions of $S_1, \ldots, S_T$ inferred from option prices, see \cite{breeden1978prices,jackwerth1996recovering}. Further, fix $f \in C_{\rm lin}(\mathbb{R}^T)$, which defines the financial instrument $f(S)$ to be priced. For arbitrary $\mathbb{Q} \in \mathcal{P}(\mathbb{R}^T)$ and a sub-tuple $I = (t_1, \ldots, t_{|I|})$ of $(1, \ldots, T)$ let $\mathbb{Q}_I := \mathbb{Q} \circ S_I^{-1}$, where $S_I: \mathbb{R}^T \rightarrow \mathbb{R}^{|I|}$ is given by $S_I(\omega) = (\omega_{t_1}, \ldots, \omega_{t_{|I|}})$. Denote by $\mathbb{Q}_t := \mathbb{Q}_{(t)}$ the $t$-th marginal of $\mathbb{Q}$, and 
\begin{align*}
\Pi(\mu_1, \ldots, \mu_T) &:= \{ \mathbb{Q} \in \mathcal{P}(\mathbb{R}^T) : \mathbb{Q}_t = \mu_t \text{ for } t = 1, \ldots, T\}, \\
\mathcal{M}(\mu_1, \ldots, \mu_T) &:= \{ \mathbb{Q} \in \Pi(\mu_1, \ldots, \mu_T): \mathbb{E}^{\mathbb{Q}}[S_{t+1}|S_1, \ldots, S_t] = S_t \text{ for all } t= 1, \ldots, T-1\}.
\end{align*}
We call $\Pi(\mu_1, \ldots, \mu_T)$ the set of all couplings between $\mu_1, \ldots, \mu_T$ and $\mathcal{M}(\mu_1, \ldots, \mu_T)$ the set of all martingale couplings.
The martingale optimal transport problem is to find the lowest and highest 
possible price of the financial instrument $f(S)$ among models in 
$\mathcal{M}(\mu_1, \ldots, \mu_T)$: Without loss of generality, we focus on the 
problem to find the highest price:
\begin{equation}
\label{eq:MOT}
\tag{MOT}
\sup_{\mathbb{Q} \in \mathcal{M}(\mu_1, \ldots, \mu_T)} ~\mathbb{E}^{\mathbb{Q}}[f(S)]
\end{equation}
In contrast, the usual (multi-marginal) optimal transport problem is stated over all couplings
\begin{equation}
\label{eq:OT}
\tag{OT}
\sup_{\mathbb{Q} \in \Pi(\mu_1, \ldots, \mu_T)} ~\mathbb{E}^{\mathbb{Q}}[f(S)].
\end{equation}
Both problems allow for a dual formulation, which can be interpreted in terms of trading. For the \eqref{eq:OT} problem, the dual formulation reads
\begin{equation}
\tag{OT-Dual}
\inf_{\substack{h_1,\dots,h_T\in C_{\rm lin}(\mathbb{R}):\\\sum_{t=1}^T h_t(S_t)\ge f(S)}} ~\sum_{t=1}^T \int_{\mathbb{R}} h_t\,d\mu_t.
\end{equation}
Here, $h_1, \ldots, h_T$ are trading strategies for a single time-period, which 
corresponds to trading freely into European call options. Indeed, one can 
restrict each $h_t$ to be a linear combination of European call options, 
i.e.,~$h_t(S_t) = \sum_{i=1}^{N} \alpha_i (S_t - k_i)^+$ for different strike 
prices $k_1, \ldots, k_N \in \mathbb{R}$, see \cite{beiglbock2013model}. For the 
\eqref{eq:MOT} problem, the martingale condition corresponds to the assumption 
that one can additionally trade dynamically in the underlying, leading to
\begin{equation}
\tag{MOT-Dual}
\inf_{\substack{h_1, \ldots, h_T\in C_{\rm lin}(\mathbb{R}),\,\vartheta_1 \in C_b(\mathbb{R}^1), \ldots, \vartheta_{T-1} \in C_b(\mathbb{R}^{T-1}) :\\\sum_{t=1}^T h_t(S_t) + \sum_{t=1}^{T-1} \vartheta_t(S_1, \ldots, S_t)\,(S_{t+1} - S_t)\ge f(S)}} ~\sum_{t=1}^T \int_{\mathbb{R}} h_t\,d\mu_t.
\end{equation}
Here, $\vartheta_t(S_1, \ldots, S_t)$ is the (positive or negative) quantity of the stock owned between times $t$ and $t+1$.

\section{Homogeneous stock movements}
\label{sec:homogeneity}
\begin{figure}[h]
	\begin{center}
		\mbox{\includegraphics[width=0.8\textwidth]{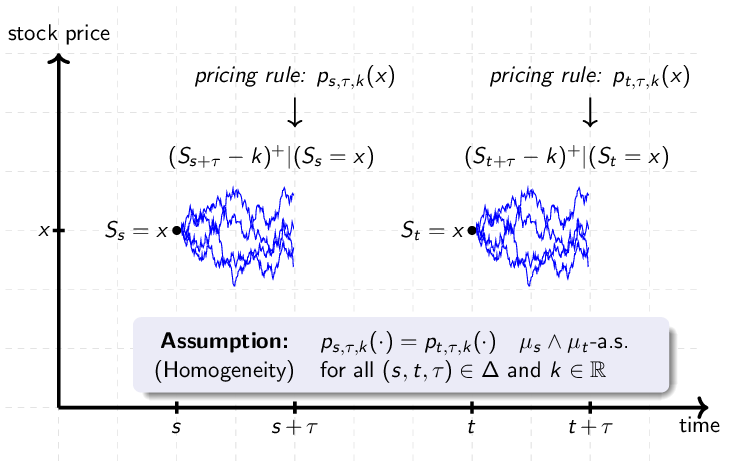}}
	\end{center}
	\caption{Illustration of Definition \ref{def:Qhom}(ii) and Remark 
		\ref{remark:basics}(i). Homogeneity for $(S_1, \ldots, S_T)$ states that the 
		forward looking option pricing rules $p_{s, \tau, k}(x)$ and $p_{t, \tau, k}(x)$ are independent of the time $s$ or $t$. For these pricing rules, we only condition on the information that the stock is in state $x$. These option pricing rules correspond to actual prices only when the considered models are Markovian. More generally they can be seen as \textsl{average} prices, averaged over possible paths that lead to state $x$.}
	\label{fig:illu}
\end{figure}

The purpose of this section is the introduction and analysis of the notion of time-homogeneity added to the martingale optimal transport setting, which restricts the variability of the forward looking transitions of the martingales across time. The formal condition is introduced in Definition \ref{def:Qhom} and illustrated in Figure \ref{fig:illu}. Basic properties are stated in Remark \ref{remark:basics}, and the duality for the time-homogeneous version of the martingale optimal transport problem is given in Theorem \ref{thm:duality}. The duality stated in Theorem \ref{thm:duality} is shortly discussed in Remark \ref{rem:interpretation} in terms of swap contracts. Remark \ref{rem:support} discusses the support of the marginals and its relation to the time-homogeneity assumption.

We first recall the following: Any $\pi \in \mathcal{P}(\mathbb{R}^2)$ can be disintegrated as $\pi = \pi_1 \otimes K$ where $\pi_1$ is the first marginal of $\pi$ and $K : \mathbb{R} \rightarrow \mathcal{P}(\mathbb{R})$ is a (Borel measurable) stochastic kernel, which is $\pi_1$-a.s.~unique. Second, for two measures $\mu, \nu$ there is a unique Lebesgue decomposition $\mu = \mu^{\nu, \rm abs} + \mu^{\nu, \rm sin}$, where $\mu^{\nu, \rm abs} \ll \nu$ and $\mu^{\nu, \rm sin} \perp \nu$. Note that $\mu^{\nu, \rm abs}$ and $\nu^{\mu, \rm abs}$ have the same null-sets.

The notation used in the following definition is fixed throughout the paper.
\begin{definition}\label{def:Qhom}
	Let $\Delta = \{(s, t, \tau) \in \{1, \ldots, T\}^3 : s<t, t+\tau \leq T\}$.
	\begin{itemize}
		\item[(i)] For $\mu, \nu \in \mathcal{P}(\mathbb{R})$ we say that an event holds $\mu \land \nu$-almost surely, if it holds almost surely with respect to $\mu^{\nu, \rm abs}$, which is the absolutely continuous part of $\mu$ with respect to $\nu$ given by Lebesgue's decomposition theorem.\footnote{This definition is consistent with the lattice minimum $\mu \land \nu$ of the two measures $\mu$ and $\nu$, see \cite[Chapter 10.10-10.11]{aliprantisborder}, which is given by $\mu \land \nu(A) := \inf_{B \subseteq A \text{ Borel}} \mu(B) + \nu(A \backslash B)$ for Borel sets $A$. One can verify that $\mu^{\nu, \rm abs}$ is equivalent to $\mu \land \nu$, see also Remark \ref{rem:naturalchoicetheta}.}
		\item[(ii)] We say that $\mathbb{Q}\in\mathcal{P}(\mathbb{R}^T)$ is homogeneous, if
		\[K_{s, s+\tau} = K_{t, t+\tau}\quad\mbox{$\mathbb{Q}_s \land \mathbb{Q}_t$-a.s. for all }(s, t, \tau)\in \Delta,\] where $K_{s, s+\tau}$ denotes the stochastic kernel given by $\mathbb{Q}_{(s, s+\tau)} = \mathbb{Q}_s \otimes K_{s, s+\tau}$.
		\item[(iii)] We set 
		\begin{align*}
		\mathcal{P}_{\rm hom}(\mathbb{R}^T) &:= \{\mathbb{Q} \in \mathcal{P}(\mathbb{R}^T) : \mathbb{Q} \text{ is homogeneous} \},\\
		\Pi_{\rm hom}(\mu_1, \ldots, \mu_T) &:= \Pi(\mu_1, \ldots, \mu_T) \cap \mathcal{P}_{\rm hom},\\
		\mathcal{M}_{\rm hom}(\mu_1, \ldots, \mu_T) &:= \mathcal{M}(\mu_1, \ldots, \mu_T) \cap \mathcal{P}_{\rm hom}.
		\end{align*}
	\end{itemize}
\end{definition}

\begin{remark}
	\label{remark:basics}
	The proof of the following statements is given in Section \ref{sec:proofs}.
	\begin{itemize}
		\item[(i)] For $\mathbb{Q} \in \Pi(\mu_1, \ldots, \mu_T)$ define the pricing rule
		\begin{align*}
		p_{s, \tau, k}(x) :=&~ \int_\mathbb{R} (y-k)^+ \,K_{s, s+\tau}(x, dy)\\
		=&~ \mathbb{E}^{\mathbb{Q}}[(S_{s+\tau}-k)^+ | S_s = x].
		\end{align*}
		Then $\mathbb{Q}$ is homogeneous if and only if $p_{s, \tau, k} = p_{t, \tau, k}$ holds $\mu_s \land \mu_t$-a.s.~for all $(s, t, \tau) \in \Delta$ and $k\in \mathbb{R}$.
		\item[(ii)] $\mathcal{P}_{\rm hom}(\mathbb{R}^T)$ is not convex, but $\Pi_{\rm hom}(\mu_1, \ldots, \mu_T)$ and $\mathcal{M}_{\rm hom}(\mu_1, \ldots, \mu_T)$ are convex and closed.
		\item[(iii)] Let $\mathcal{P}_{\rm HM}(\mathbb{R}^T)$ be the set of measures $\mathbb{Q} \in \mathcal{P}(\mathbb{R}^T)$ such that the canonical process $(S_1, \ldots, S_T)$ is a homogeneous Markov chain under $\mathbb{Q}$. It holds $\conv(\mathcal{P}_{\rm HM}(\mathbb{R}^T)) \subset \mathcal{P}_{\rm hom}(\mathbb{R}^T)$, which is strict for $T \geq 3$.
		\item[(iv)] It holds $\Pi_{\rm hom}(\mu_1, \ldots, \mu_T)\neq \emptyset$ if and only if $(\mu_1, \ldots, \mu_{T-1})$ dominates $(\mu_2, \ldots, \mu_T)$ in heterogeneity (see \cite[Definition 3.4.]{shen2019distributional}).\footnote{We thank Ruodu Wang for pointing this relation out to us.} We state a simplified definition of domination in heterogeneity for convenience: Let $\kappa := \frac{1}{T} \sum_{t=1}^T \mu_t$, then $(\mu_1, \ldots, \mu_{T-1})$ dominates $(\mu_2, \ldots, \mu_T)$ in heterogeneity if for all convex functions $\varphi: \mathbb{R}^{T-1} \rightarrow \mathbb{R}$ it holds $\int \varphi\big(\frac{d\mu_1}{d\kappa}, \ldots, \frac{d\mu_{T-1}}{d\kappa}\big) \,d\kappa \geq \int \varphi\big(\frac{d\mu_2}{d\kappa}, \ldots, \frac{d\mu_{T}}{d\kappa}\big) \,d\kappa$. We refer to the introduction of \cite{shen2019distributional} and Chapter 9.7.~of \cite{torgersen1991comparison} for more details regarding domination in heterogeneity.
		\item[(v)] A condition for $\mathcal{M}_{\rm hom}(\mu_1, ..., \mu_T) \neq \emptyset$ is unknown to us at the moment, and appears difficult to obtain. Notably, it is not the case that 
		\[(\mathcal{M}_{\hom}(\mu_1, \ldots, \mu_T) \neq \emptyset) \Leftrightarrow (\Pi_{hom}(\mu_1, \ldots, \mu_T) \neq \emptyset \land \mathcal{M}(\mu_1, \ldots, \mu_T) \neq \emptyset).\]
	\end{itemize}
\end{remark}


\noindent
To state duality, we make the following assumption: 
\begin{description}
	\item[$(A)$] For all $(s, t) \in \{1, \ldots, T\}^2$ there exists a finite Borel measure $\theta^{s, t}$ on $\mathbb{R}$ which is equivalent to $\mu_t^{\mu_s, \rm abs}$ such that $\frac{d\theta^{s, t}}{d\mu_t}$ and $\frac{d\theta^{s, t}}{d\mu_s}$ are continuous and bounded.
\end{description}
This is satisfied in a large number of cases, for example if all marginals $\mu_1, \ldots, \mu_t$ are discrete, or if all marginals have a continuous and strictly positive Lebesgue density. In fact, the latter generalizes to the case where $\mu_1, \ldots, \mu_T$ all have a continuous and strictly positive density with respect to any reference measure $\theta$. Then, one can define $\theta^{s, t}$ by $\frac{d\theta^{s, t}}{d\theta} := \min\{\frac{d\mu_t}{d\theta}, \frac{d\mu_s}{d\theta}\}$ and finds that $\frac{d\theta^{s, t}}{d\mu_t}$ and $\frac{d\theta^{s, t}}{d\mu_s}$ are continuous and bounded by 1. We further note that properties (like continuity) of densities are always understood in the sense that there exists a representative among the almost sure equivalence class satisfying the property. For the statement of the theorem, one such representative satisfying this property is then fixed.

Even though Assumption $(A)$ is quite general, there are examples which do not satisfy it, see Example \ref{ex:counterA}. Necessity of this assumption is hinted at in Remark \ref{rem:naturalchoicetheta}.

We now state the main result of the paper.

\begin{theorem}\label{thm:duality} Let $\mu_1, \ldots, \mu_T \in \mathcal{P}(\mathbb{R})$ have finite first moment, and $f \in C_{\rm lin}(\mathbb{R}^T)$. Assume $\mathcal{M}_{\rm hom}(\mu_1, \ldots, \mu_T) \neq \emptyset$ and $(A)$ holds. Then
	\begin{align*}
	\tag{HMOT}\label{eq:duality}
	&\hspace{-10mm}\max_{\mathbb{Q}\in\mathcal{M}_{\rm hom}(\mu_1,\dots,\mu_T)}\mathbb{E}^{\mathbb{Q}}[f(S)]\\
	= \inf\bigg\{ &\sum_{t=1}^T \int_{\mathbb{R}} h_t \, d\mu_t : ~~~ &&h_t \in C_{\rm lin}(\mathbb{R}), ~t\in\{1, \ldots, T\},\\
	& &&\vartheta_t \in C_b(\mathbb{R}^t), ~t\in\{1, \ldots, T-1\},\\
	& &&g_{s, t, \tau} \in C_b(\mathbb{R}^2), ~(s, t, \tau) \in \Delta,\\
	& \text{ such that } && f(S) \leq \sum_{t=1}^T h_t(S_t)+\sum_{t=1}^{T-1}\vartheta_{t}(S_1,\dots,S_t)\big(S_{t+1}-S_t\big) \\
	& && + \sum_{(s, t, \tau) \in \Delta} \Big(g_{s, t, \tau}(S_s, S_{s+\tau}) \frac{d\theta^{s, t}}{d\mu_s}(S_s) - g_{s, t, \tau}(S_t, S_{t+\tau}) \frac{d\theta^{s, t}}{d\mu_t}(S_t)\Big)\bigg\}
	\end{align*}
\end{theorem}
The proof of Theorem \ref{thm:duality} is given in Section \ref{sec:proofs}.
\begin{remark}
	\label{rem:interpretation}
	Compared to the usual martingale optimal transport, the additional trading term arising from the homogeneity condition in the dual formulation is the sum
	\[
	\sum_{(s, t, \tau) \in \Delta} \Big(g_{s, t, \tau}(S_s, S_{s+\tau}) \frac{d\theta^{s, t}}{d\mu_s}(S_s) - g_{s, t, \tau}(S_t, S_{t+\tau}) \frac{d\theta^{s, t}}{d\mu_t}(S_t)\Big).
	\]
	Each individual summand can be interpreted as a swap contract which, under the assumption of time-homogeneity, has fair price 0 from today's point of view. In Markovian models, the terms can even be hedged dynamically, as the homogeneous pricing rules correspond to actual prices on the market.
	
	To simplify, say $g_{s, t, \tau}(S_s, S_{s+\tau}) = V(S_s) \cdot 
	(S_{s+\tau}-k)^+$. 
	This means, one buys $V(S_s)$ many call options at time $s$ which expire at time $s+\tau$. Under homogeneity, see Remark \ref{remark:basics} (i), conditioned on $S_s = x$, the expected price from today's point of view of such a financial instrument is the same when replacing time point $s$ with some other time point $t$. If two traders were to agree that such an instrument is equally valuable for time points $s$ and $t$, it has to be taken into consideration how likely the events $S_s = x$ and $S_t = x$ are. The fair weighting to take this into account is achieved by the terms $\frac{d\theta^{s, t}}{d\mu_s}(S_s)$ and $\frac{d\theta^{s, t}}{d\mu_t}(S_t)$. In Markovian models, when conditioning on the events $S_s = x$ or $S_t=x$, these instruments are not just equally valuable from today's point of view, but the actual prices on the market given these states are the same as well.
\end{remark}

\begin{remark}
	\label{rem:support}
	The assumption of homogeneity crucially depends on the respective support of the marginals. Indeed, if $\mu_t$ and $\mu_s$ have disjoint support, the condition
	\[K_{s, s+\tau} = K_{t, t+\tau}\quad\mbox{$\mu_s \land \mu_t$-a.s.}\]
	is empty. For practical purposes, one might want to strengthen the condition so that it is robust with respect to slight perturbations of the marginal supports. For instance, if marginal $\mu_s$ has support $\{1, 2\}$ and $\mu_t$ has support $\{1-\varepsilon, 2+\varepsilon\}$, a strengthening of homogeneity might require $K_{s, s+\tau}(1) \approx K_{t, t+\tau}(1-\varepsilon)$. While such a strengthening is intuitive and sensible for discrete supports, it is more difficult to formalize in full generality. Nevertheless, such an approximate equality between stochastic kernels appears related to concepts like nested distance (see for instance \cite{backhoff2019adapted} and references therein).
\end{remark}

\section{Extensions}
\label{sec:extensions}
This section aims at discussing the following extensions and variations of the approach:
\begin{enumerate}
	\item Non-equally spaced time-grids
	\item Variations of the time-homogeneity assumption
	\item Market frictions
	\item Extension to several assets (high-dimensional market)
\end{enumerate}
The first two points are specific to the setting at hand, while the latter two points are reoccurring themes in robust pricing. We hence go briefly over the latter two issues, while referencing related work.
\paragraph{Non-equally spaced time-grids.} The notion of time-homogeneity introduced in Section \ref{sec:homogeneity} makes sense when subsequent time steps are equally far apart, which is the case if there exists a constant $C$ such that time points $t$ and $s$ are $|t-s|\cdot C$ many trading days apart. Available data on option prices is not always equally spaced. The framework can account for this by modeling the time steps as $t_1 < t_2 < \ldots < t_N$ with $t_i \in \mathbb{N}$ (instead of $t=1, \ldots, T$), where $|t_i - t_j|$ measures the number of trading days between time points $t_i$ and $t_j$. Then one can set
\[
\tilde\Delta := \{(i, j, \tau_i, \tau_j) \in \{1, \ldots, N\}^4 : i < j, i+\tau_i \leq N, j+\tau_j \leq N, |t_{i+\tau_i} - t_i| = |t_{j+\tau_j}-t_j|\}
\]
and Definition \ref{def:Qhom} (i) changes to $K_{i, \tau_i} = K_{j, \tau_j}$ for all $(i, j, \tau_i, \tau_j) \in \tilde\Delta$, where now $\mathbb{Q}_{(t_i, t_{i+\tau_i})} = \mu_{t_i} \otimes K_{i, \tau_i}$, etc. If for the available option maturities the set $\tilde\Delta$ is not large enough, or even empty, one can consider relaxing the equality constraint $|t_{i+\tau_i} - t_i| = |t_{j+\tau_j}-t_j|$ to $||t_{i+\tau_i} - t_i| - |t_{j+\tau_j}-t_j|| \leq C$ for some constant $C > 0$. As one now couples transition probabilities for time intervals of possibly different length, this can be combined with weakening the notion of time-homogeneity, see below.

\paragraph{Variations of the time-homogeneity assumption.} A natural stronger version of time-homogeneity is the extension from one-period transitions to many-period transitions. For two-period transitions for instance, Definition \ref{def:Qhom} can be extended via the condition 
\[K_{s, (s+\tau_1, s+\tau_2)} = K_{t, (t+\tau_1, t+\tau_2)} \quad\mbox{$\mathbb{Q}_s \land \mathbb{Q}_t$-a.s. for all }(s, t, \tau_1), (s, t, \tau_2)\in \Delta \text{ with } \tau_1 < \tau_2,\]
where $\mathbb{Q}_{(s, s+\tau_1, s+\tau_2)} = \mathbb{Q}_s \otimes K_{s, (s+\tau_1, s+\tau_2)}$ and $K_{s, (s+\tau_1, s+\tau_2)} : \mathbb{R} \rightarrow \mathcal{P}(\mathbb{R}^2)$. With such an extension, all relevant properties like the convexity of $\Pi_{\rm hom}(\mu_1, \ldots, \mu_T)$ remain unchanged.

Weakening the notion of time-homogeneity can be done in various ways. First, note that 
\[K_{s, \tau} = K_{t, \tau}  ~~ \mu_s \land \mu_t \text{-a.s.} ~~ \Leftrightarrow ~~ \theta^{s, t} \otimes K_{s, \tau} = \theta^{s, t} \otimes K_{t, \tau}\]
with $\theta^{s, t}$ as in condition $(A)$ stated before Theorem \ref{thm:duality}. So time-homogeneity can simply be stated as equalities of measures. A natural relaxation is to instead assume that the measures are close in a suitable distance $D(\cdot, \cdot)$, like Wasserstein-distance or relative entropy. The relaxation from homogeneity to $r$-homogeneity takes the form
\[
\theta^{s, t} \otimes K_{s, \tau} = \theta^{s, t} \otimes K_{t, \tau} ~~\stackrel{\text{Relaxation}}{\longrightarrow}~~ D(\theta^{s, t} \otimes K_{s, \tau}, \theta^{s, t} \otimes K_{t, \tau}) \leq r_{s, t, \tau},
\]
where $r_{s, t, \tau} \geq 0$ for all $(s, t, \tau) \in \Delta$.
If the mapping $(\mu, \nu) \mapsto D(\mu, \nu)$ is convex, the set of $r$-homogeneous couplings between $\mu_1, \ldots, \mu_T$ remains convex.

Alternatively, one can directly penalize the distance between $\theta^{s, t} \otimes K_{s, \tau}$ and $\theta^{s, t} \otimes K_{t, \tau}$ in the statement of the optimization problem, which leads to
\[
\label{eq:Pen-HMOT}
\tag{Pen-HMOT}
\sup_{\mathbb{Q} \in \mathcal{M}(\mu_1, \ldots, \mu_T)} \mathbb{E}^{\mathbb{Q}}[f(S)] - \sum_{(s, t, \tau) \in \Delta} \frac{1}{r_{s, t, \tau}} D(\theta^{s, t} \otimes K_{s, \tau}, \theta^{s, t} \otimes K_{t, \tau}).
\]
For appropriately chosen $D(\cdot, \cdot)$, this penalization corresponds to the inclusion of transaction costs in the dual formulation, which is discussed below.

\paragraph{Market frictions.}
The most flexible notion of market frictions that can be incorporated in the framework is that of transaction costs. Transaction costs result in more costly hedging strategies on the dual side, and in relaxed constraints for the considered models $\mathbb{Q}$ on the primal side. Proportional transaction costs correspond to an enlargement of the set of feasible models, see e.g.~\cite{cheridito2017duality,dolinsky2014robust,guo2019computational}. With superlinear transaction costs on the other hand, the constraint is completely removed, and instead a penalization term is added to the objective function, see e.g.~\cite{bank2016super,cheridito2017duality}. 

An instance of such a penalized primal formulation resulting from superlinear 
transaction costs is the above defined \eqref{eq:Pen-HMOT} for appropriately 
chosen penalization term. If $D = G$ is the Gini index in 
\eqref{eq:Pen-HMOT},\footnote{For two measures $\nu, \mu$ the Gini index $G$ is 
	defined as $G(\nu, \mu) = \int \Big(\frac{d\nu}{d\mu}\Big)^2 d\mu-1$, if $\nu 
	\ll \mu$ and $G(\nu, \mu) = \infty$, else. See also 
	\cite{maccheroni2004variational}.} this corresponds to the use of quadratic 
transaction costs in the dual formulation, which means the term
\[
g_{s, t, \tau}(S_s, S_{s+\tau}) \frac{d\theta^{s, t}}{d\mu_s}(S_s) - g_{s, t, \tau}(S_t, S_{t+\tau}) \frac{d\theta^{s, t}}{d\mu_t}(S_t)
\]
would incur transaction costs
\[
4\,r_{s, t, \tau}\,|g_{s, t, \tau}(S_s, S_{s+\tau})|^2\frac{d\theta^{s, t}}{d\mu_s}(S_s).
\]
So it holds
\begin{corollary}
	Under the assumptions of Theorem \ref{thm:duality}, it holds
	\begin{align*}\tag{$G$-Pen-HMOT}\label{eq:gpenhmot}
	&\hspace{-10mm}\max_{\mathbb{Q} \in \mathcal{M}(\mu_1, \ldots, \mu_T)} \mathbb{E}^{\mathbb{Q}}[f(S)]&&\hspace{-3mm} - \sum_{(s, t, \tau) \in \Delta} \frac{1}{r_{s, t, \tau}} G(\theta^{s, t} \otimes K_{s, \tau}, \theta^{s, t} \otimes K_{t, \tau})\\
	= \inf\bigg\{ &\sum_{t=1}^T \int_{\mathbb{R}} h_t \, d\mu_t : ~~~ &&h_t \in C_{\rm lin}(\mathbb{R}), ~t\in\{1, \ldots, T\},\\
	& &&\vartheta_t \in C_b(\mathbb{R}^t), ~t\in\{1, \ldots, T-1\},\\
	& &&g_{s, t, \tau} \in C_b(\mathbb{R}^2), ~(s, t, \tau) \in \Delta,\\
	& \text{ such that } && f(S) \leq \sum_{t=1}^T h_t(S_t)+\sum_{t=}^{T-1}\vartheta_{t}(S_1,\dots,S_t)\big(S_{t+1}-S_t\big) \\
	& && + \sum_{(s, t, \tau) \in \Delta} \Big(g_{s, t, \tau}(S_s, S_{s+\tau}) \frac{d\theta^{s, t}}{d\mu_s}(S_s) - g_{s, t, \tau}(S_t, S_{t+\tau}) \frac{d\theta^{s, t}}{d\mu_t}(S_t) \\
	& && \hspace{21mm} -4\,r_{s, t, \tau}\,|g_{s, t, \tau}(S_s, S_{s+\tau})|^2\frac{d\theta^{s, t}}{d\mu_s}(S_s)\Big)\bigg\}
	\end{align*}
\end{corollary}
The proof is sketched in Section \ref{subsec:prooftransaction}.
In general, the time-homogeneity assumption behaves quite similarly to the martingale or marginal assumptions in terms of transaction costs, and hence many different modeling approaches can be applied.

In one respect, the notion of time-homogeneity is however more restrictive: 
When including the assumption of time-homogeneity, one has to take care with 
relaxing the assumption of precisely knowing the marginal laws. The reason is 
that $\mathcal{P}_{\hom}(\mathbb{R}^T)$ is not convex, and the optimization 
problem only becomes feasible by adding constraints so that the resulting set 
of models $\mathbb{Q}$ is convex (this is achieved by specifying marginal 
distributions, see Remark \ref{remark:basics}(ii)), which is crucial for the 
tractability of the resulting optimization problem.


\paragraph{Extension to several assets.} The martingale optimal transport 
setting generalizes as follows to higher dimensions: One specifies $\mu_{t, i}$ 
of each stock $S_{t, i}$ for time points $t=1, \ldots, T$ and dimensions $i=1, 
\ldots, d$ individually. The dimensions are coupled through a joint martingale 
constraint $\mathbb{E}[S_{t+1, i}|S_1,\ldots,S_t] = S_{t, i}$ where $S_t = (S_{t, 
	1}, \ldots, S_{t, d})$.\footnote{See also \cite{jgts2019mmmot, lim2016multi}. Some 
	papers \cite{de2019irreducible,ghoussoub2019structure,obloj2017structure} 
	extend the MOT problem in a different way, where the assumption is made that for each 
	time point, the $d$-dimensional marginal distribution is known. While it leads 
	to an interesting mathematical problem, this assumption is less well justified 
	from a financial viewpoint, as one can only infer each individual 
	one-dimensional marginal distribution from market data.} 

In terms of homogeneity, one has two choices: First, one can define the notion 
of homogeneity as in Definition \ref{def:Qhom} (ii) for each dimension $i = 1, 
\ldots, d$ individually. This is straightforward and sensible, and the resulting 
optimization problem remains convex. An alternative to take into consideration 
is to specify homogeneity jointly across dimensions, similarly to the 
martingale constraint. Then, Definition \ref{def:Qhom} is stated for 
$\mathbb{Q}\in \mathcal{P}((\mathbb{R}^d)^T)$, and $\mathbb{Q}_t \in 
\mathcal{P}(\mathbb{R}^d)$, etc. In this case however, convexity of the set of 
models becomes an issue. If only the individual one-dimensional 
marginals of $S_{t, i}$ are known, the set of
time-homogeneous multi-dimensional martingale optimal transport measures will 
not be convex.\footnote{Intuitively, the same reason as for $\mathcal{P}_{\rm hom}(\mathbb{R}^T)$ applies, see Remark \ref{remark:basics} (ii). If the complete marginals at a time point are not fixed, taking the convex combinations of joint distributions no longer corresponds to convex combinations of transition kernels.} Hence, for mathematical purposes, the first alternative is more 
suitable.

\section{Examples}
In this section, we present two short numerical examples that showcase the potential of the introduced setting.

\subsection{Discrete model}
\label{subsec:binomial}
\begin{figure}
	\begin{minipage}{0.5\textwidth}
		\begin{center}
			\mbox{\includegraphics[width=1\textwidth]{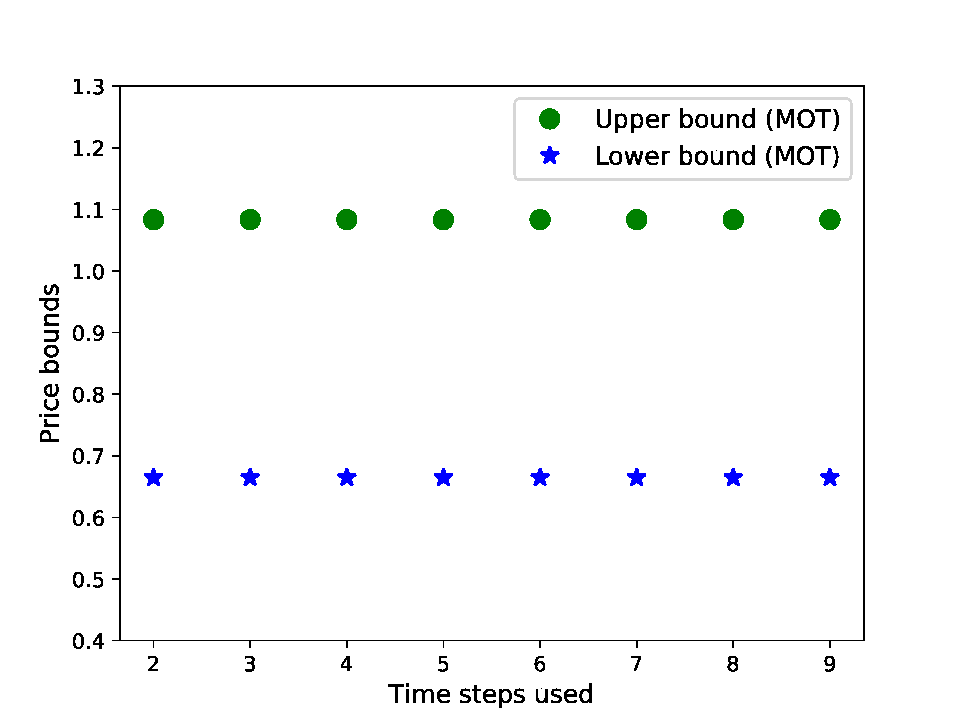}}
		\end{center}
	\end{minipage}
	\begin{minipage}{0.5\textwidth}
		\begin{center}
			\mbox{\includegraphics[width=1\textwidth]{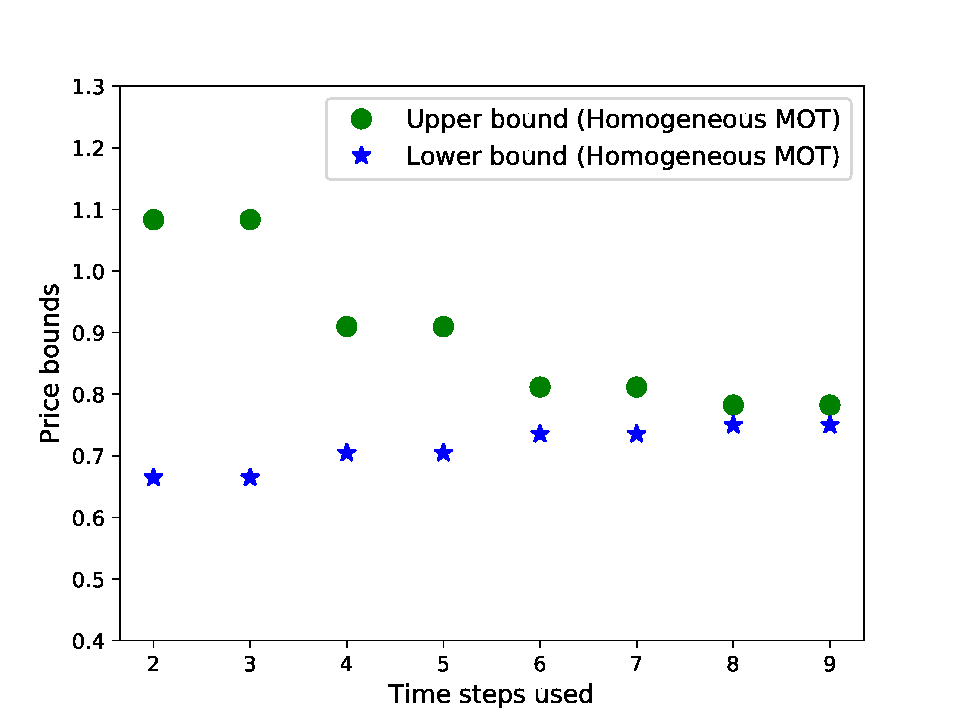}}
		\end{center}
	\end{minipage}
	\caption{Discrete example from Subsection \ref{subsec:binomial} illustrated. Price bounds for a financial instrument $f(S) = (S_{9}-S_8)^+$ are depicted. For both figures, time steps used indicates how many marginal distributions are known, i.e., how much market data is used. We see that for the martingale optimal transport approach alone, using more data does not improve the obtained price bounds. Incorporating homogeneity however leads to improved bounds when adding data. 
	}
	\label{fig:binomial}
\end{figure}
Consider a discrete model where $\mu_t$ is the uniform distribution on the set $\{100-t, 100-t+2, \ldots, 100+t\}$ for $t=1, \ldots, 9$.\footnote{So for $t=0$ the set is $\{100\}$, for $t=1$ the set is $\{99, 101\}$, for $t=2$ the set is $\{98, 100, 102\}$, and so on.} So the support of the marginals is the same as in a binomial model where the stock starts at $100$ at time point $0$ and can either go up or down by $1$ each period. The financial instrument is a forward start option, $f(S) = (S_{9}-S_8)^+$. First, we solve the model using just the data (i.e., marginal distributions) from time points $t=8, 9$ (two time steps used). Then, we gradually increase the information that is used, by adding the marginal information from $t=7$ (three time steps used), $t=6$, etc.~until all marginals $\mu_1, \ldots, \mu_{9}$ are included (nine time steps used). The results are reported in Figure \ref{fig:binomial}. On the left, we see that without the homogeneity assumption, the bounds do not get sharper with inclusion of additional information. With the added assumption of homogeneity however, the bounds tighten drastically. As seen in Figure \ref{fig:binomial}, the bounds only tighten once every second increment of additional information. The reason for this is the support of the marginals: Indeed, the only transition probability that actually matters for pricing the financial instrument $f$ is the one from $t=8$ to $t=9$. So the relevant kernel lives on the support of $\mu_8$. Hence only by adding marginal information $\mu_t$ which shares support with $\mu_8$, the bounds tighten, see also Remark \ref{rem:support}.
While in this example the bounds merely tighten drastically, there is no guarantee in general that including many time steps coupled with the homogeneity assumption does not lead to infeasibility.

\subsection{Black-Scholes model}
Let $T=3$ and $\mu_t \sim X_t$ for $t=1, 2, 3$, where $X_t = X_0 \exp(\sigma W_t - \frac{\sigma^2}{2}t)$ for $\sigma > 0$ is a geometric Brownian motion. Following \cite{alfonsi2019Journal,sester2018markov} we consider the option $f(S_1, S_2, S_3) := (S_3 - \frac{S_1 + S_2}{2})^+$. Set $X_0 = 1$, $\sigma = 0.25$. The model price for the Black-Scholes model $\mathbb{Q}_{\rm BS}$ is given by
\[
\mathbb{E}^{\mathbb{Q}_{\rm BS}}[f(S)] \approx 0.111.
\]
Compared to the previous example, where the (homogeneous) martingale transport problem is a linear program, the current example has to be solved approximately. Discretization is non-trivial even with just the martingale condition, see \cite{alfonsi2019Journal,guo2019computational}. Homogeneity, which crucially depends on the given marginals' support, adds difficulty for a discretization scheme. Hence, we instead calculate this example using the dual formulation and the penalization approach of \cite{eckstein2018computation}, i.e., we approximate each trading strategy $h_t, \vartheta_t$ and $g_{s, t, \tau}$ by a neural network.\footnote{To approximate a trading strategy with $d$ inputs, we use a network structure with 5 layers, hidden dimension $64 \cdot d$ and ReLu activation function. For the penalization as in \cite{eckstein2018computation}, we use the product measure $\theta = \mu_1 \times \mu_2 \times \mu_3$ and $\beta_{\gamma}(x) = 10000 \max\{0, x\}^2$. For training, we use batch size 8192, learning rate $0.0001$ (after the first 60000 iterations, learning rate is decreased by a factor of 0.98 each 250 iterations for another 60000 iterations), and the Adam optimizer with default parameters. The reported values are primal values, as described at the start of Section 4 in \cite{eckstein2018computation}.} Without the homogeneity, this leads to
\[
\inf_{\mathbb{Q} \in \mathcal{M}(\mu_1, \mu_2, \mu_3)} \mathbb{E}^{\mathbb{Q}}[f(S)] \approx 0.059 ~~~ \text{ and } ~~~ \sup_{\mathbb{Q} \in \mathcal{M}(\mu_1, \mu_2, \mu_3)} \mathbb{E}^{\mathbb{Q}}[f(S)] \approx 0.139.\footnote{In comparison, using the discretization scheme from \cite[Subsection 6.4.]{alfonsi2019Journal} with 100 samples for each marginal, we get $\inf_{\mathbb{Q} \in \mathcal{M}(\mu_1, \mu_2, \mu_3)} \mathbb{E}^{\mathbb{Q}}[f(S)] \approx 0.058$ and $\sup_{\mathbb{Q} \in \mathcal{M}(\mu_1, \mu_2, \mu_3)} \mathbb{E}^{\mathbb{Q}}[f(S)] \approx 0.139$.}
\]
On the other hand, incorporating homogeneity improves the bounds slightly but notably to
\[
\inf_{\mathbb{Q} \in \mathcal{M}_{\rm hom}(\mu_1, \mu_2, \mu_3)} \mathbb{E}^{\mathbb{Q}}[f(S)] \approx 0.064 ~~~ \text{ and } ~~~ \sup_{\mathbb{Q} \in \mathcal{M}_{\rm hom}(\mu_1, \mu_2, \mu_3)} \mathbb{E}^{\mathbb{Q}}[f(S)] \approx 0.135.
\]
While the strengths of the homogeneity assumption certainly lie with cases where more time steps are involved, even in this example the bounds are narrowed by around $11\%$.

Assuming homogeneity of the underlying process also becomes more restrictive when the marginals $\mu_1, \mu_2, \mu_3$ are less homogeneously evolving. As an extreme case, if in the above we instead set $\mu_3 \sim X_4$, then the interval of possible prices for the MOT is $[0.088, 0.184]$ while for the homogeneous MOT one obtains $[0.121, 0.138]$, which is drastically more narrow.
\section{Proofs}
\label{sec:proofs}
\subsection{Proof of Remark \ref{remark:basics}}
\textsl{Proof of (i):} If $\mathbb{Q}$ is homogeneous, then by definition $p_{s, \tau, k} = p_{t, \tau, k}$ holds $\mu_s \land \mu_t$-a.s.. The reverse follows since the function class $\{h(x) = (x-k)^+ : k \in \mathbb{R}\}$ is measure determining, see e.g.~\cite[Footnote 2]{beiglbock2013model}.

\vspace{2mm}
\noindent
\textsl{Proof of (ii):} First, we show that $\mathcal{P}_{\rm hom}(\mathbb{R}^T)$ is not convex. Consider $T=3$ and the two homogeneous Markov chains $\mathbb{Q}^a = 0.75 \,\delta_{(0, 1, 0)} + 0.25 \,\delta_{(1, 0, 1)}$ and $\mathbb{Q}^b = 0.75\,\delta_{(0, 0, 0)} + 0.25\, \delta_{(1, 1, 1)}$. Both Markov chains start in state $0$ with probability $0.75$ and state 1 with probability $0.25$. Chain $a$ always switches states and chain $b$ always stays in the same state. Obviously $\mathbb{Q}^a, \mathbb{Q}^b \in \mathcal{P}_{\rm hom}(\mathbb{R}^3)$. But $\mathbb{Q} := 0.5 \mathbb{Q}^a + 0.5 \mathbb{Q}^b \not\in\mathcal{P}_{\rm hom}(\mathbb{R}^3)$. Indeed, at time $1$ the Markov chain transitions from state $0$ to each state with equal probability. With the notation as in Definition \ref{def:Qhom}, it holds $K_{1, 2}(0) = 0.5 \,\delta_0 + 0.5 \, \delta_1$. However, at time $2$ one gets $K_{2, 3}(0) = 0.75\, \delta_0 + 0.25 \,\delta_1$.

Next, we show that $\Pi_{\rm hom}(\mu_1, \ldots, \mu_T)$ is convex, which implies that $\mathcal{M}_{\rm hom}(\mu_1, \ldots, \mu_T)$ is convex too. Let $\mathbb{Q}^a, \mathbb{Q}^b \in \Pi_{\rm hom}(\mu_1, \ldots, \mu_T)$, $\lambda \in (0, 1)$ and $\mathbb{Q} := \lambda \mathbb{Q}^a + (1-\lambda) \mathbb{Q}^b$. Take $(s, t, \tau) \in \Delta$. We have to show $K_{s, s+\tau} = K_{t, t+\tau}$. With notation as in Definition \ref{def:Qhom}, $\mathbb{Q}_{(s, s+\tau)} = \mathbb{Q}_s\otimes K_{s, s+\tau}$. Denote by $K^a_{s, s+\tau}$ the stochastic kernel satisfying $\mathbb{Q}^a_{(s, s+\tau)} = \mu_s \otimes K^a_{s, s+\tau}$ (same for $K^b_{s, s+\tau}$). By the general formula
\[
K_{(s, s+\tau)} = \lambda \frac{d \mathbb{Q}^a_s}{d\mathbb{Q}_s} K^a_{s, s+\tau} + (1-\lambda) \frac{d \mathbb{Q}^b_s}{d\mathbb{Q}_s} K^b_{s, s+\tau},
\]
and since all measures have the same marginals, it follows $K_{s, s+\tau} = \lambda K^a_{s, s+\tau} + (1-\lambda) K^b_{s, s+\tau}$, which yields the claim.

Finally, we show that $\Pi_{\rm hom}(\mu_1, ..., \mu_T)$ is closed, which implies that $\mathcal{M}_{\rm hom}(\mu_1, \ldots, \mu_T)$ is closed too.\footnote{We thank one of the reviewers for pointing this proof out to us.}
Choose $\theta^{s, t} := \mu_t \land \mu_s$.
Let $\pi^n \in \Pi_{\rm 
	hom}(\mu_1, \ldots, \mu_T)$ with $\pi^n \stackrel{w}{\rightarrow} \pi$. For $\pi \in 
\Pi_{\rm hom}(\mu_1, \ldots, \mu_T)$ we have to show $\theta^{s, t} \otimes 
K_{s, s+\tau} = \theta^{s, t} \otimes K_{t, t+\tau}$ where $\pi_{(t, 
	t+\tau)} = \mu_t \otimes K_{t, t+\tau}$ as usual. We further use the 
notation $\pi^n_{(t, t+\tau)} = \mu_t \otimes K^n_{t, t+\tau}$. Then it 
holds for $\varphi \in C_b(\mathbb{R}^2)$
\begin{align*}
\int \varphi \,d\theta^{s, t}\otimes K_{s, s+\tau} &= \int \varphi(x, y) 
\frac{d\theta^{s, t}}{d\mu_s}(x) \,\pi_{(s, s+\tau)}(dx, dy) \\
&\stackrel{(*)}{=} \lim_{n\rightarrow \infty} \int \varphi(x, y) \frac{d\theta^{s, 
		t}}{d\mu_s}(x) 
\,\pi^n_{(s, s+\tau)}(dx, dy) \\
&= \lim_{n\rightarrow \infty} \int \varphi \,d\theta^{s, t}\otimes K^n_{s, s+\tau} \\
&= \lim_{n\rightarrow \infty} \int \varphi \,d\theta^{s, t}\otimes K^n_{t, t+\tau} \\
&= \ldots \\
&= \int \varphi \,d\theta^{s, t}\otimes K_{t, t+\tau},
\end{align*}
where $\ldots$ are the same steps as above but reversed for $t$ instead of $s$. 
The step $(*)$ follows by weak convergence and since the first marginals of 
$\pi^n$ and $\pi$ are fixed. This allows an approximation of $\frac{d\theta^{s, 
		t}}{d\mu_s}$ by continuous and bounded functions via Lusin's theorem which 
coincides with $\frac{d\theta^{s, t}}{d\mu_s}$ on compacts with measure almost 
1 under $\mu_s$, and both $\pi$ and $\pi^n$ have first marginal $\mu_s$.

\vspace{2mm}
\noindent
\textsl{Proof of (iii):} The inclusion is trivial: Indeed, if $K : \mathbb{R} \rightarrow \mathcal{P}(\mathbb{R})$ is the transition kernel of the homogeneous Markov chain, then with the notation as in Definition \ref{def:Qhom}, it holds $K_{s, s+\tau} = K^{\tau}$, which is independent of $s$. (Hereby, $K^s$ is defined as usual by $K^{s+1}(x, A) := \int K(y, A) K^s(x, dy)$.)

That the inclusion is strict, consider the following example: Let $\mathbb{Q}^a := 0.5\,\delta_{(0, 0, 1)} + 0.5\,\delta_{(1, 1, 0)}$ and $\mathbb{Q}^b := 0.5\,\delta_{(0, 1, 1)} + 0.5\,\delta_{(1, 0, 0)}$. Intuitively, $\mathbb{Q}^a$ stays constant after the first period, and switches states after the second period, and $\mathbb{Q}^b$ does exactly the reverse. In particular, the corresponding processes are not Markovian. It holds $\mathbb{Q} := 0.5 \mathbb{Q}^a + 0.5\mathbb{Q}^b \in \mathcal{P}_{\rm hom}(\mathbb{R}^3)$. However, straightforward calculation shows that $\mathbb{Q}$ cannot be written as a convex combination of homogeneous Markov chains, so $\mathbb{Q} \not\in \conv(\mathcal{P}_{\rm HM}(\mathbb{R}^3))$.

\vspace{2mm}
\noindent
\textsl{Proof of (iv):} Define $\Pi_{\rm HM}(\mu_1, \ldots, \mu_T) := \Pi(\mu_1, \ldots, \mu_T) \cap \mathcal{P}_{\rm HM}(\mathbb{R}^T)$. 

In a first step we show that $\Pi_{\rm hom}(\mu_1, \ldots, \mu_T) \neq \emptyset$ if and only if $\Pi_{\rm HM}(\mu_1, \ldots, \mu_T) \neq \emptyset$. Indeed, by inclusion, the 'if' direction is clear. On the other hand, for $\mathbb{Q} \in \Pi_{\rm hom}(\mu_1, \ldots, \mu_T)$ we can define $\mathbb{Q}^{\rm HM} \in \Pi_{\rm HM}(\mu_1, \ldots, \mu_T)$ as follows: We use the notation $\mathbb{Q}_{(t, t+1)} = \mu_t \otimes K_{t, t+1}$ and show the following statement inductively over $t=2, \ldots, T$: There exists a coupling $\mathbb{Q}^{\rm HM}_{(1, \ldots, t)} \in \Pi_{\rm HM}(\mu_1, \ldots, \mu_t)$ with $\mathbb{Q}^{\rm HM}_{(1, \ldots, t)} = \mu_1 \otimes S \otimes \ldots \otimes S$, where $S(x) = K_{s, s+1}(x)$ holds for $\mu_s$ almost all $x \in \mathbb{R}$ for all $s=1, \ldots, t-1$. For $t=2$, this clearly holds. Assume we have such a coupling for $t$, and we now construct $\mathbb{Q}^{\rm HM}_{(1, \ldots, t+1)} \in \Pi_{\rm HM}(\mu_1, \ldots, \mu_{t+1})$ with the same property. To this end, we first note that for all $s=1, \ldots, t-1$ we can find Borel sets $\Omega_{s, 1}, \Omega_{s, 2}, \Omega_{s, 3} \subset \mathbb{R}$ with $\Omega_{s, 1} \stackrel{.}{\cup} \Omega_{s, 2} \stackrel{.}{\cup} \Omega_{s, 3} = \mathbb{R}$ and
\begin{align*}
\mu_{s}^{\mu_{t},  \rm abs}(\Omega_{s, 1}) &= \mu_{s}^{\mu_{t},  \rm abs}(\mathbb{R}), \\
\mu_{s}^{\mu_{t},  \rm sin}(\Omega_{s, 2}) &= \mu_{s}^{\mu_{t},  \rm sin}(\mathbb{R}), \\
\mu_{t}^{\mu_{s},  \rm sin}(\Omega_{s, 3}) &= \mu_{t}^{\mu_{s},  \rm sin}(\mathbb{R}).
\end{align*}
Further, without loss of generality assume that (by changing kernels on a null set)
\begin{itemize}
	\item on $\Omega_{s, 1}$ it holds $K_{s, s+1} = K_{t, t+1}$ by homogeneity,
	\item and on $\Omega_{s, 1} \cup \Omega_{s, 2}$ it holds $S = K_{s, s+1}$.
\end{itemize}
The set $\Omega_2 := \cup_{s=1}^{t-1} \Omega_{s, 2}$ is a $\mu_t$ null-set and $\Omega^3 := \cap_{s=1}^{t-1} \Omega_{s, 3}$ is a $\mu_s$ null-set for all $s=1, \ldots, t-1$. Now, define $\tilde{S} : \mathbb{R} \rightarrow \mathcal{P}(\mathbb{R})$ by
\[
\tilde{S}(x) := \left\{\begin{array}{ll}
K_{t, t+1}(x),& x \in \Omega_3, \\
S(x),& \text{else}.
\end{array}\right.
\]
Then, $\tilde{S}(x) = K_{s, s+1}(x)$ holds $\mu_s$ almost surely, since $\Omega_3$ is a $\mu_s$ null-set for $s=1, \ldots, t-1$. Further, $\tilde{S}(x) = K_{t, t+1}(x)$ also holds $\mu_t$ almost surely: If $x \not\in \Omega_3$, then either $x \in \Omega_2$, which is a $\mu_{t}$ null-set, or $x \in \Omega_1 := \cup_{s=1}^{t-1} \Omega_{s, 1}$. And if $x \in \Omega_1$, then there exists some $s\in\{1, \ldots, t-1\}$ such that $K_{t, t+1}(x) = K_{s, s+1}(x)$ by homogeneity, and hence $K_{t, t+1}(x) = K_{s, s+1}(x) = S(x) = \tilde{S}(x)$ by induction. We get $\mathbb{Q}^{\rm HM}_{(1, \ldots, t+1)} := \mu_1 \otimes \tilde{S} \otimes \ldots \otimes \tilde{S} \in \Pi_{\rm HM}(\mu_1, \ldots, \mu_{t+1})$ with $\tilde{S}(x) = K_{s, s+1}(x)$ for $\mu_s$ almost all $x \in \mathbb{R}$ for all $s=1, \ldots, t$.

To complete the proof, by \cite[Theorem 9.7.3, (iii) $\Leftrightarrow$ (iv')]{torgersen1991comparison} it follows that $\Pi_{\rm HM}(\mu_1, \ldots, \mu_T) \neq \emptyset$ if and only if $(\mu_1, \ldots, \mu_{T-1})$ dominates $(\mu_2, \ldots, \mu_T)$ in heterogeneity.

\textsl{Proof of (v):} The simple counterexample to the stated equivalence is $\mu_1 = \delta_0$ and $\mu_2 = \mu_3 = \frac{1}{3}(\delta_{-1} + \delta_0 + \delta_1)$, since there exists only one martingale coupling, which is not homogeneous, and the coupling which always spreads its mass equally to each point is a homogeneous coupling.
\qed

\subsection{Proof of Theorem \ref{thm:duality}}
Define $\phi(f)$ as the infimum term in the statement of the theorem for $f\in C_{\rm lin}(\mathbb{R}^T)$. For $\pi \in \mathcal{P}(\mathbb{R}^T)$, the convex conjugate $\phi^\ast$ is given by
\[
\phi^*(\pi) := \sup_{f\in C_{\rm lin}(\mathbb{R}^T)} \Big( \int f \,d\pi - \phi(f)\Big).
\]
We show $\phi(f) = \sup_{\pi \in \mathcal{P}(\mathbb{R}^T)} \int f \,d\pi - \phi^\ast(\pi)$ using \cite[Theorem 2.2.]{bartl2019robust} and calculate $\phi^\ast(\pi)$ so that the proposition follows.

\textsl{Dual representation:} To apply \cite[Theorem 2.2.]{bartl2019robust}, we show that $\phi(f)$ is real-valued on $C_{\rm lin}(\mathbb{R}^T)$ and condition (R1) stated within the Theorem holds, i.e., for $C_{\rm lin}(\mathbb{R}^T) \ni f_n \downarrow 0$ it holds $\phi(f_n) \downarrow \phi(0)$ for $n\rightarrow \infty$. Regarding $\phi(f) \in \mathbb{R}$, $\phi(f) < \infty$ is obvious. On the other hand, $\phi(f) > -\infty$ will follow by calculation of $\phi^\ast(\pi)$ and the assumption that $\mathcal{M}_{\rm hom}(\mu_1, \ldots, \mu_T)$ is non-empty, since for $\pi \in \mathcal{M}_{\rm hom}(\mu_1, \ldots, \mu_T)$ then $0 = \phi^\ast(\pi) \geq \int f \,d\pi - \phi(f)$ and since $\int f \,d\pi < \infty$ (all marginals have first moments and $f \in C_{\rm lin}(\mathbb{R}^T)$), it holds $\phi(f) > -\infty$. Regarding condition (R1), note that $\phi(0) = 0$ and $\phi(f) \leq \phi_{\rm ot}(f)$ where $\phi_{\rm ot}$ is the optimal transport functional \[\phi_{\rm ot}(f) := \inf_{\substack{h_1, \ldots, h_T \in C_{\rm lin}(\mathbb{R}):\\\forall x \in \mathbb{R}^T: \sum_{t=1}^T h_t(x_t) \geq f(x)}} \sum_{t=1}^T \int_\mathbb{R} h_t \,d\mu_t.\] Since $\phi_{\rm ot}$ is continuous from above on $C_{\rm lin}(\mathbb{R}^T)$ (see \cite[Proof of Theorem 1]{eckstein2018robust}), $\phi$ is as well.

\textsl{Computation of the convex conjugate:} We show $\phi^*(\pi) = 0$, if 
$\pi \in \mathcal{M}_{\rm hom}(\mu_1, \ldots, \mu_T)$, and $\phi^*(\pi) = \infty$, 
else. After plugging in the definitions and exchanging suprema, one obtains
\[
\phi^{\ast}(\pi) = \sup_{h_t, \vartheta_t, g_{s, t, \tau}} \sup_{\substack{f \in C_{\rm lin}(\mathbb{R}^T):\\f \leq T(h_t, \vartheta_t, g_{s, t, \tau})}} \int f \,d\pi - \sum_{t=1}^T \int_\mathbb{R} h_t \,d\mu_t
\]
where $T(h_t, \vartheta_t, g_{s, t, \tau}) \in C_{\rm lin}(\mathbb{R}^T)$ is the term
\begin{align*}
T(h_t, \vartheta_t, g_{s, t, \tau}) = \sum_{t=1}^T h_t(S_t)+\sum_{t=1}^{T-1}\vartheta_{t}(S_1,\dots,S_t)\big(S_{t+1}-S_t\big)\\ + \sum_{(s, t, \tau) \in \Delta} \Big(g_{s, t, \tau}(S_s, S_{s+\tau}) \frac{d\theta^{s, t}}{d\mu_s}(S_s) - g_{s, t, \tau}(S_t, S_{t+\tau}) \frac{d\theta^{s, t}}{d\mu_t}(S_t)\Big).
\end{align*}
Hence the inner supremum is attained for $f = T(h_t, \vartheta_t, g_{s, t, \tau})$.
It follows:
\begin{align*}
\phi^*(\pi) &= \sup_{h_t \in C_{\rm lin}(\mathbb{R})} \sum_{t=1}^T \int_{\mathbb{R}} h_t\,d\pi_t - \int_{\mathbb{R}} h_t \,d\mu_t \tag{a} \\
&+ \sup_{\vartheta_t \in C_b(\mathbb{R}^t)} \sum_{t=1}^{T-1} \int_{\mathbb{R}^T} \vartheta_t(x_1, \ldots, x_t)\cdot(x_{t+1}-x_t) \,\pi(dx_1, \ldots, dx_T) \tag{b} \\
\tag{c}
\begin{split}
&+ \sup_{g_{s, t, \tau} \in C_b(\mathbb{R}^2)} \sum_{(s, t, \tau)\in \Delta} \int_{\mathbb{R}^T} \Big(g_{s, t, \tau}(x_s, x_{s+\tau}) \frac{d\theta^{s, t}}{d\mu_s}(x_s) \\
&\hspace{4.5cm}- g_{s, t, \tau}(x_t, x_{t+\tau}) \frac{d\theta^{s, t}}{d\mu_t}(x_t)\Big) \,\pi(dx_1, \ldots, dx_T)
\end{split}
\end{align*}
By martingale optimal transport duality, we have: Term (a) is zero if $\pi_t = \mu_t$ for all $t=1, \ldots, T$, and else infinity. Term (b) is zero if the canonical process is a martingale under $\pi$, and else infinity. It only remains to show that term (c) is zero if $\pi$ is homogeneous, and else infinity. This is done already under the assumption that $\pi_t = \mu_t$ for all $t=1, \ldots, T$. We write $\pi_{(t, t+\tau)} = \pi_t \otimes K_{t, t+\tau}$. Then one calculates for $(s, t, \tau) \in \Delta$ and $g_{s, t, \tau} \in C_b(\mathbb{R}^2)$
\begin{align*}
&\int_{\mathbb{R}^T} \Big(g_{s, t, \tau}(x_s, x_{s+\tau}) \frac{d\theta^{s, t}}{d\mu_s}(x_s) - g_{s, t, \tau}(x_t, x_{t+\tau}) \frac{d\theta^{s, t}}{d\mu_t}(x_t)\Big) \,\pi(dx_1, \ldots, dx_T) \\
= &\int_{\mathbb{R}^2} g_{s, t, \tau}(x_s, x_{s+\tau}) \frac{d\theta^{s, t}}{d\mu_s}(x_s) \,\pi_{(s, s+\tau)}(dx_s, dx_{s+\tau}) \\
&-\int_{\mathbb{R}^2} g_{s, t, \tau}(x_t, x_{t+\tau}) \frac{d\theta^{s, t}}{d\mu_t}(x_t) \,\pi_{(t, t+\tau)}(dx_t, dx_{t+\tau}) \\
=& \int_{\mathbb{R}^2} g_{s, t, \tau}(x_s, x_{s+\tau}) \,\theta^{s, t} \otimes K_{s, s+\tau}(dx_s, dx_{s+\tau}) -\int_{\mathbb{R}^2} g_{s, t, \tau}(x_t, x_{t+\tau}) \,\theta^{s, t} \otimes K_{t, t+\tau}(dx_t, dx_{t+\tau}) \\
=& \int_{\mathbb{R}^2}g_{s, t, \tau} \,d\theta^{s, t}\otimes K_{s, s+\tau} - \int_{\mathbb{R}^2} g_{s, t, \tau} \,d\theta^{s, t}\otimes K_{t, t+\tau}
\end{align*}
And hence, term (c) is zero if $\theta^{s, t} \otimes K_{s, s+\tau} = \theta^{s, t}\otimes K_{t, t+\tau}$ for all $(s, t, \tau) \in \Delta$, and else infinity. So it is zero if and only if $K_{t, t+\tau} = K_{s, s+\tau}$ holds $\theta^{s, t}$-a.s. for all $(s, t, \tau) \in \Delta$. This, by choice of $\theta^{s, t}$, corresponds to $\mu_s \land \mu_t$-a.s.~equality, and hence term (c) is zero if and only if $\pi$ is homogeneous.\qed
\subsection{Inclusion of transaction costs}
\label{subsec:prooftransaction}
We sketch the proof for the duality formula shown for \eqref{eq:gpenhmot} from Section \ref{sec:extensions}. The argument works the same as for Theorem \ref{thm:duality}. The only difference is term (c), which for the case of transaction costs reads
\begin{align*}
\tag{c}
\begin{split}
\sup_{g_{s, t, \tau} \in C_b(\mathbb{R}^2)} \sum_{(s, t, \tau)\in \Delta} \int_{\mathbb{R}^T} \Big(&g_{s, t, \tau}(x_s, x_{s+\tau}) \frac{d\theta^{s, t}}{d\mu_s}(x_s)- g_{s, t, \tau}(x_t, x_{t+\tau}) \frac{d\theta^{s, t}}{d\mu_t}(x_t)\\
&-4\,r_{s, t, \tau}\,|g_{s, t, \tau}(x_s, x_{s+\tau})|^2\frac{d\theta^{s, t}}{d\mu_s}(x_s)
\Big) \,\pi(dx_1, \ldots, dx_T).
\end{split}
\end{align*}
Similar to the proof of Theorem \ref{thm:duality}, using $\pi_t = \mu_t$ for $t=1, \ldots, T$, this simplifies to
\begin{align*}
\sum_{(s, t, \tau)\in \Delta} \sup_{g_{s, t, \tau} \in C_b(\mathbb{R}^2)}  &\int_{\mathbb{R}^2} g_{s, t, \tau} \,d\theta^{s, t}\otimes K_{s, s+\tau} - \int_{\mathbb{R}^2} g_{s, t, \tau} \,d\theta^{s, t}\otimes K_{t, t+\tau} \\&- 4r_{s, t, \tau} \int_{\mathbb{R}^2}|g_{s, t, \tau}|^2 \,d\theta^{s, t}\otimes K_{s, s+\tau}
\end{align*}
and finally the terms inside the sum are the dual representation for the Gini index $G(\theta^{s, t}\otimes K_{s, s+\tau}, \theta^{s, t}\otimes K_{t, t+\tau})$ as shown in \cite{maccheroni2004variational} (in \cite{maccheroni2004variational} the space $L^2$ instead of $C_b$ is used, but since continuous and bounded functions are dense in $L^p$, the above representation follows), which yields the claim.

\appendix 
\section{Discussion of Assumption $(A)$}
We shortly discuss Assumption $(A)$. In particular, we showcase that it is non-trivial (meaning there are cases where it is not satisfied) in Example \ref{ex:counterA}, and in Remark \ref{rem:naturalchoicetheta} we give indications that the assumption might be necessary for duality. 

\begin{example}
	\label{ex:counterA}
	Let $Q_i = \{q_{i, 1}, q_{i, 2}, q_{i, 3}, ...\}$ for $i=1, 2, 3$ be three disjoint, countable, dense subsets of $[0, 1]$. Let $\kappa_i := \sum_{j=1}^\infty 2^{-j} \delta_{q_{i, j}}$ for $i=1, 2, 3$ and $\mu_1 := \frac{1}{2} (\kappa_1 + \kappa_2)$, $\mu_2 := \frac{1}{2} (\kappa_2 + \kappa_3)$. Then Assumption $(A)$ is not satisfied for $\mu_1$ and $\mu_2$. Indeed, the measure $\mu_1^{\mu_2, \rm abs}$ is given by $\kappa_2$, and hence any measure $\theta^{1, 2}$ which has the same null sets has to have support $Q_2$. Hence $\frac{d\theta^{1, 2}}{d\mu_1}$ is strictly positive on $Q_2$, but it is zero on $Q_1$, and hence not continuous.
\end{example}

\begin{remark}
	\label{rem:naturalchoicetheta}
	\begin{itemize}
		\item[(i)] By definition, the null sets that $\theta^{s, t}$ should represent for Assumption $(A)$ are given by both $\mu_{t}^{\mu_s, \rm{abs}}$ and $\mu_{s}^{\mu_t, \rm{abs}}$. However, another natural choice for $\theta^{s, t}$ is the lattice minimum $\mu_t \land \mu_s$, see \cite[Chapter 10.10-10.11]{aliprantisborder}. The definition of the lattice minimum $\mu_t \land \mu_s$ is given by \[\mu_t \land \mu_s (A) = \inf_{B \subseteq A \text{ Borel }} \mu_{t}(B) + \mu_s(A\backslash B)\] which is equivalent to  \[\mu_t \land \mu_s (A) = \inf_{B \subseteq A \text{ Borel }} \mu_{t}^{\mu_s, \rm{abs}}(B) + \mu_s^{\mu_t, \rm{abs}}(A\backslash B)\] for Borel sets $A \subseteq \mathbb{R}$. Taking $\theta^{s, t} = \mu_t \land \mu_s$, both $\frac{\mu_t \land \mu_s}{d\mu_t}$ and $\frac{\mu_t \land \mu_s}{d\mu_s}$ are bounded by 1.
		\item[(ii)] Part (i) shows that without the continuity part of Assumption $(A)$, the assumption would always be satisfied by choosing $\theta^{s, t} = \mu_t \land \mu_s$. 
		
		However, without the continuity condition, Theorem \ref{thm:duality} does in general not hold: Consider a case where all marginals are equivalent to the Lebesgue measure. For any choice of $\theta^{s, t}$, one can pick representatives within the equivalence classes of the relevant densities which are equal to 0 on a dense subset of $\mathbb{R}$. On this dense set, the term in the hedging inequality arising from homogeneous trading, which is
		\[
		\sum_{(s, t, \tau) \in \Delta} \Big(g_{s, t, \tau}(S_s, S_{s+\tau}) \frac{d\theta^{s, t}}{d\mu_s}(S_s) - g_{s, t, \tau}(S_t, S_{t+\tau}) \frac{d\theta^{s, t}}{d\mu_t}(S_t)\Big),
		\]
		is equal to 0. Thus the problem $\phi(f)$ reduces to pure martingale 
		optimal transport, since by continuity this dense set determines the 
		pointwise hedging term. 
		
		This showcases that one requires some smoothness condition on the respective densities. The continuity condition as currently given in Assumption $(A)$ is sufficient, but may not be the weakest possible assumption in this respect.
	\end{itemize}
\end{remark}

\footnotesize
\bibliographystyle{abbrv}
\bibliography{HomBib}

\end{document}